\documentclass{pasj01}
\usepackage{bm}
\usepackage{natbib,aas_macros}
\usepackage{comment}
\usepackage{color}
\usepackage{txfonts}
\usepackage{ulem}

\Received{}
\Accepted{}
\newcommand{\simgt}{\lower.5ex\hbox{$\; \buildrel > \over \sim \;$}}
\newcommand{\simlt}{\lower.5ex\hbox{$\; \buildrel < \over \sim \;$}}

\newcommand{\red}{\textcolor{red}}

\definecolor{blue-violet}{rgb}{0.54, 0.17, 0.89}
\definecolor{bostonuniversityred}{rgb}{0.8, 0.0, 0.0}
\definecolor{brilliantrose}{rgb}{1.0, 0.33, 0.64}
\definecolor{calpolypomonagreen}{rgb}{0.12, 0.3, 0.17}
\definecolor{auburn}{rgb}{0.43, 0.21, 0.1}
\definecolor{brightpink}{rgb}{1.0, 0.0, 0.5}
\definecolor{amber}{rgb}{1.0, 0.49, 0.0}
\definecolor{bananayellow}{rgb}{1.0, 0.88, 0.21}
\definecolor{capri}{rgb}{0.0, 0.75, 1.0}

\newcommand{\norv}[1]{{\textcolor{blue}{#1}}}

\def\h70kpc{\mathrel{h_{70}^{-1}{\rm kpc}}}

\def\h70Msol{\mathrel{h_{70}^{-1}M_\odot}}
 
\begin{document} 

\title{ 
Signatures of large-scale cold fronts in the optically-selected merging cluster HSC J085024+001536 }



\author{Keigo \textsc{Tanaka}\altaffilmark{1}%
\thanks{Example: Present Address is xxxxxxxxxx}}
\altaffiltext{1}{Graduate School of Natural Science \& Technology, Kanazawa University, Kakuma-machi, Kanazawa, Ishikawa 920-1192, Japan}
\email{tanaka@astro.s.kanazawa-u.ac.jp}

\author{Ryuichi \textsc{Fujimoto}\altaffilmark{2}}
\altaffiltext{2}{Faculty of Mathematics and Physics, Kanazawa University, Kakuma-machi, Kanazawa, Ishikawa 920-1192, Japan}

\author{Nobuhiro \textsc{Okabe}\altaffilmark{3,4,5}}
\altaffiltext{3}{Physics Program, Graduate School of Advanced Science and Engineering, Hiroshima University, 1-3-1 Kagamiyama, Higashi-Hiroshima, Hiroshima 739-8526, Japan}
\altaffiltext{4}{Hiroshima Astrophysical Science Center, Hiroshima University, 1-3-1 Kagamiyama, Higashi-Hiroshima, Hiroshima 739-8526, Japan}
\altaffiltext{5}{Core Research for Energetic Universe, Hiroshima University, 1-3-1, Kagamiyama, Higashi-Hiroshima, Hiroshima 739-8526, Japan}

\author{Ikuyuki \textsc{Mitsuishi}\altaffilmark{6}}
\altaffiltext{6}{Department of Physics, Nagoya University, Furo-cho, Chikusa-ku, Nagoya, Aichi 464-8602, Japan}

\author{Hiroki \textsc{Akamatsu}\altaffilmark{7}}
\altaffiltext{7}{SRON Netherlands Institute for Space Research, Sorbonnelaan 2, 3584 CA Utrecht, the Netherlands}

\author{Naomi \textsc{Ota}\altaffilmark{8}}
\altaffiltext{8}{Department of Physics, Nara Women's University, Kitauoyanishi-machi, Nara, 630-8506, Japan}

\author{Masamune \textsc{Oguri}\altaffilmark{9,10,11}}
\altaffiltext{9}{Department of Physics, University of Tokyo, Tokyo 113-0033, Japan}
\altaffiltext{10}{Institute of Astronomy, The University of Tokyo, 2-21-1 Osawa, Mitaka, Tokyo 181-0015, Japan}
\altaffiltext{11}{Kavli Institute for the Physics and Mathematics of the Universe (Kavli IPMU, WPI), University
of Tokyo, Chiba 277-8582, Japan}

\author{Atsushi J. \textsc{Nishizawa}\altaffilmark{12}}
\altaffiltext{12}{Institute for Advanced Research, Nagoya University Furocho, Chikusa-ku, Nagoya, 464-8602 Japan}



\maketitle

\begin{abstract}
We represent a joint X-ray, weak-lensing, and optical analysis of the optically-selected merging cluster, HSC J085024+001536, from the Subaru HSC-SSP survey. Both the member galaxy density and the weak-lensing mass map show that the cluster is composed of southeast and northwest components. 
The two-dimensional weak-lensing analysis shows that the southeast component is the main cluster, and the sub- and main-cluster mass ratio is $0.32^{+0.75}_{-0.23}$. 
The northwest subcluster is offset by $\sim700$ kpc from the main cluster center, and their relative line-of-sight velocity is $\sim1300\,{\rm km s^{-1}}$ from spectroscopic redshifts of member galaxies. 
The X-ray emission is concentrated around the main cluster, while the gas mass fraction within a sphere of $1'$ radius of the subcluster is only  $f_{\mathrm{gas}}=4.0^{+2.3}_{-3.3}\%$, indicating that 
the subcluster gas was stripped by ram pressure. X-ray residual image shows three arc-like excess patterns, of which two are symmetrically located at $\sim550$~kpc from the X-ray morphological center, and the other is close to the X-ray core. The excess close to the subcluster has a cold-front feature where dense-cold gas and thin-hot gas contact. The two outer excesses are tangentially elongated about $\sim 450-650$ kpc, suggesting that the cluster is merged with a non-zero impact parameter. Overall features revealed by the multi-wavelength datasets indicate that the cluster is at the second impact or later. Since the optically-defined merger catalog is unbiased for merger boost of the intracluster medium, X-ray follow-up observations will pave the way to understand merger physics at various phases.

\end{abstract}

\section{Introduction}

Cluster mergers are the most energetic phenomena in the Universe.
Along with the large energy release, cluster mergers have a significant impact on both the thermal history and cluster evolution \citep[e.g.,][]{Ricker01,ZuHone11} through shock waves induced by supersonic motions of subclusters.
Subsonic motions trigger perturbations in the intracluster medium (ICM) distribution of main clusters and mix 
the two different ICM of the main and sub clusters  \citep[e.g.,][]{2006ApJ...650..102A,ZuHone10}. 
The feature of the mixing is sometimes observed as cold fronts \citep[e.g.,][]{2007PhR...443....1M,Zuhone16}. 
The merging process can be broadly divided into three phases of pre-merger, on-going, and post-merger by a sound-crossing time scale.
The ICM temperature and X-ray luminosity increase around the on-going phase, so-called merger boost, and subsequently decrease at the post-merger phase because of adiabatic expansion. 
According to \citet{Ricker01} simulations, the temperature and the X-ray luminosity increment could be by a factor of about four and ten, respectively, for a head-on equal mass merger, and the time scale over which luminosity and temperature are boosted is of the order of the sound-crossing time.
The subcluster gradually decelerates before reaching a turn-around distance, and then it starts the second impact.
The increase of X-ray luminosity due to the merger boost at the second core passage is much smaller than that at the first one, only $30-70\%$ for a head-on equal mass merger \citep{Ricker01}. The consecutive time scale of cluster mergers is an order of gigayear and much shorter than the dynamical time. 
Due to the merger boost effect, a certain probability of finding merging clusters through X-ray observations is biased toward the on-going phase.
Therefore, most previous studies focus on on-going mergers, while studies of pre-/post-phase merging clusters are very limited. In order to understand the overall process of cluster mergers, it is vitally important to study cluster mergers at various merger phases using a homogeneous merging cluster sample with a selection function independent of the merger boost of the ICM \citep{Okabe2019}.



The Hyper Suprime-Cam Subaru Strategic Program
\citep[HSC-SSP;][]{HSC1stDR,HSC1styrOverview,Miyazaki18HSC,Komiyama18HSC,Kawanomoto18HSC,Furusawa18HSC,Bosch18HSC,Haung18HSC,Coupon18HSC,HSC2ndDR}
is an on-going wide-field optical imaging survey composed of three layers of different depths (Wide, Deep, and UltraDeep). 
The Wide layer is designed to obtain five-band ($grizy$) imaging over $1400$~deg$^2$.
The HSC-SSP Survey achieves both excellent imaging quality ($\sim$0.7 arcsec
seeing in $i$-band) and deep observations ($r\simlt26$~AB~mag).
Therefore, the HSC-SSP Survey currently has the best performance to search for
galaxy clusters and to measure their weak-lensing masses \citep[for review;][]{Pratt19}.
\citet{Oguri18} constructed a cluster catalogue using the {\it Cluster finding Algorithm based
on Multi-band Identification of Red-sequence gAlaxies} \citep[CAMIRA;][]{Oguri14b}.
The catalogue contains $\sim 1900 $ clusters at $0.1<z<1.1$ with richness
larger than $N=15$ in the $\sim 240$~deg$^2$ HSC-SSP S16A field.
\citet{Okabe2019} found $\sim 190$ major-merger candidates using a peak-finding method
of galaxy maps of the CAMIRA clusters and confirmed that the subcluster mass is more than one-tenth of the main cluster mass from the stacked weak-lensing analysis. The galaxy distribution provides us with unique and ideal information to construct a homogeneous sample of major mergers because distributions of long-lifetime galaxy subhalos are similar to those of dark matter subhalo \citep{Okabe08,Okabe14a}. The sample can cover from pre- to post- mergers in various dynamical stages. 
However, it is difficult to distinguish merging phases only from optical data. Therefore, X-ray follow-up observations of optically selected merging clusters are critical to identify cluster merger phases as well as to study cluster merger physics, especially at the non-on-going phase. 

HSC J085024+001536 is one of the major-merger candidates \citep{Okabe2019} at (RA, DEC)=(\timeform{8h50m23.9s}, \timeform{0D15'36.4''}) (J2000). The richness of cluster members is $N\sim 60$, and its photometric redshift is $z=0.1966$ \citep{Oguri18}. The brightest cluster galaxy (BCG) in HSC J085024+001536 corresponds to SDSS J085027.77+001501.5, of which redshift is $z=0.19608$ \citep{Driver2011}.


In this paper, we present the analysis results of HSC J085024+001536,
as the first object of joint X-ray, optical, and weak-lensing study
of optically selected merging clusters from the HSC-SSP Survey.
Using optical data, we investigate galaxy density distribution of HSC J085024+001536 
(\S\ref{subsec:galden}), construct a weak-lensing mass map,
and measure its weak-lensing mass (\S\ref{sec:WL}).
We then search for any residual structures in X-ray images (\S\ref{subsec:ximg}), 
and any unique temperature structures by X-ray spectral analysis (\S\ref{sec:spec}). 
Based on these results, we discuss the spatial distribution of the ICM and the merging history (\S\ref{sec:discuss}).

Throughout this paper, we adopt a Hubble constant of $H_0=70$\,km\,s$^{-1}$\,Mpc$^{-1}$, and cosmological density parameters of $\Omega_{\rm m,0}=0.27$ and $\Omega_{\Lambda,0}=0.73$. In this cosmology, $1'$ corresponds to 196 kpc at the redshift of $z=0.1966$. All error ranges are 68\% confidence intervals unless otherwise stated.

\section{Data analysis and results}
\label{sec:analysis}
\subsection{Galaxy density distribution}
\label{subsec:galden}

\begin{figure}
    \begin{center}
    \includegraphics[width=8.0cm]{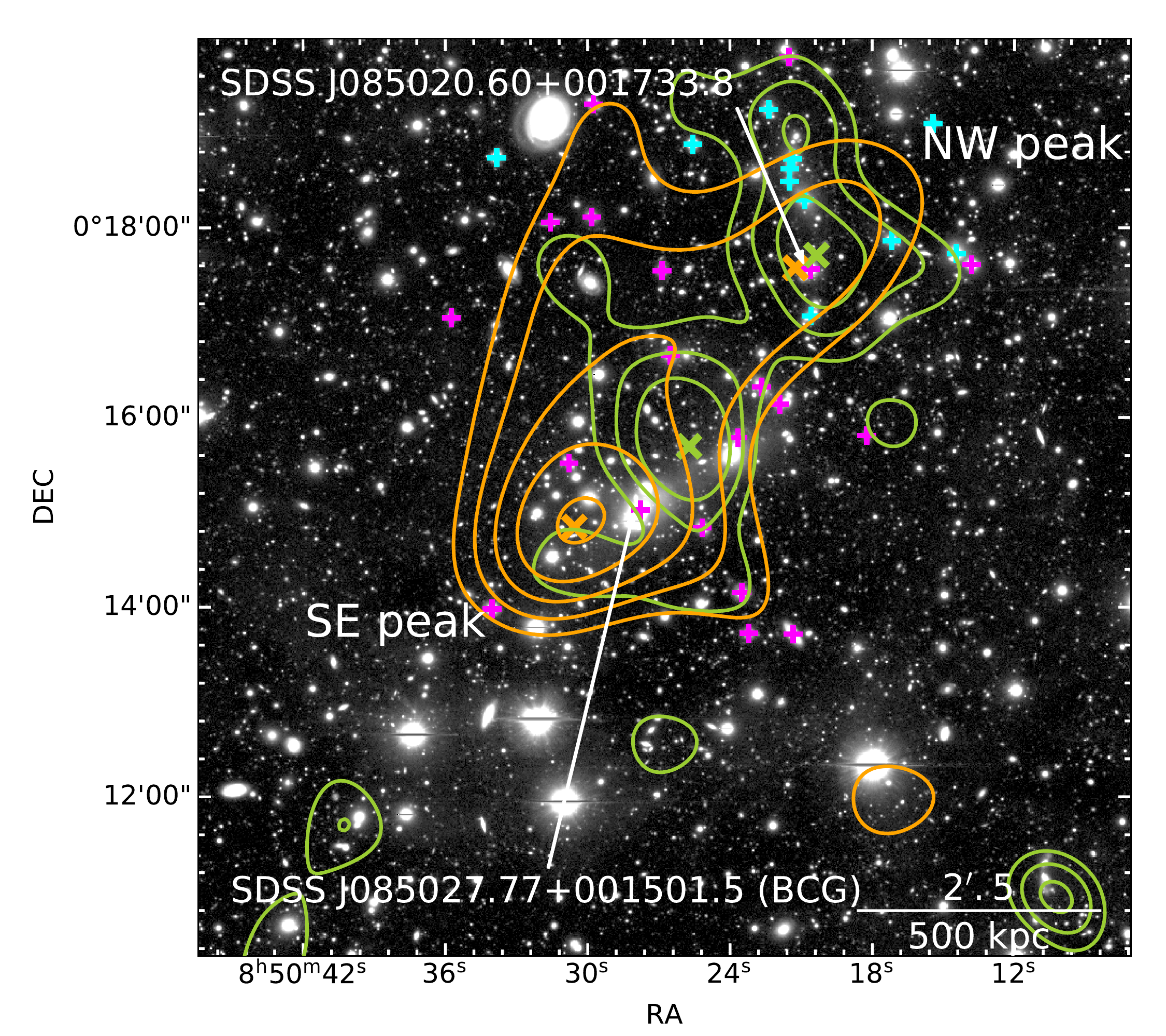}
    \end{center}
    \caption{Contours of the galaxy density distribution (green) and of the weak-lensing mass map 
    (orange), superposed on the optical image observed with Subaru/HSC. 
    The green and orange x-marks show the peaks of
    the galaxy density and the weak-lensing mass, respectively. 
    Contours of the galaxy density and the 
    weak-lensing mass are smoothed with $\sigma=200$ and 400 kpc, respectively. 
    The sky-blue and magenta $+$ signs show distribution of galaxies of which redshifts are $0.188<z<0.193$ and $0.193<z<0.201$, respectively, taken from the GAMA DR3 spectroscopic redshift catalogue \citep{GAMA_DR3}. 
    }
    \label{fig:opt-gal-WL}
\end{figure}

We first make galaxy density distribution using the Subaru HSC-SSP S19A data, i.e., the data obtained till the first semester of 2019 \citep{HSC2ndDR}. We selected red-sequence galaxies of which $z$-band magnitude is brighter than 24 ABmag in the color-magnitude plane, following \citet{Nishizawa2018} and \citet{Okabe2019}. We adopt the smoothing scale of 200$h_{70}^{-1}$ kpc. The green contours in figure~\ref{fig:opt-gal-WL} show the galaxy density distribution overlaid on a Subaru-HSC i-band image. The galaxy density distribution has two peaks, the northwest peak and the southeast peak (hereafter NW peak and SE peak, respectively). 
The NW peak corresponds to the galaxy SDSS J085020.60+001733.8.
The SE peak is offset to the northwest from the brightest cluster galaxy (BCG; SDSS J085027.77+001501.5) because the BCG is too bright for the Subaru telescope and the BCG and its surrounding galaxies are missed by photometry flags. 

Figure \ref{fig:opt-gal-WL} also shows the distribution of galaxies that have a spectroscopic redshift of $0.188<z<0.201$ in the GAMA DR3 catalogue \citep{GAMA_DR3}.
Redshifts of the galaxies around the SE peak and NW peak are $z\sim 0.197$ and $z\sim0.191$, respectively.

\subsection{Weak-lensing analysis}
\label{sec:WL}


We use the re-Gaussianization method \citep{Hirata03}, which is implemented in the HSC pipeline 
(see details in \cite{HSCWL1styr}) to measure galaxy shapes.  We select only galaxies satisfying the full-color and full-depth criteria from the HSC galaxy catalogue for both our precise shape measurements and photometric redshift estimations.

We compute weak-lensing (WL) mass reconstruction, so-called mass map, following \cite{Okabe08}. The mass map does not assume any mass models and thus is complementary to the model fitting. We adopt the FWHM$=400h_{70}^{-1}$~kpc smoothing scale. Figure \ref{fig:opt-gal-WL} compares the galaxy density and a weak-lensing mass map. 
The weak-lensing mass map elongates from northwest to southeast. The main, SE peak coincides with the BCG position within the smoothing scale. The NW weak-lensing peak is associated with the NW peak of the galaxy density distribution.

We carry out an NFW model \citep{NFW96} fitting with a free central position
using a two-dimensional shear pattern \citep{Oguri10b,Okabe11} to measure the mass of the main and subhalos. 
The formulation of weak-lensing mass measurements is described in detail in the Appendix.
The three-dimensional mass density profile of the NFW profile \citep{NFW96} is
expressed as, 
\begin{equation}
\rho_{\rm NFW}(r)=\frac{\rho_s}{(r/r_s)(1+r/r_s)^2},
\label{eq:rho_nfw}
\end{equation}
where $r_s$ is the scale radius, and $\rho_s$ is the central density
parameter. The NFW model is also specified by the spherical mass,
$M_{\Delta}=4\pi \Delta \rho_{\rm cr} r_{\Delta}^3/3$, and the halo
concentration, $c_{\Delta}=r_{\Delta}/r_s$. Here, $r_\Delta$ is the overdensity radius, and we use $\Delta=500$. 
Table \ref{tab:WL_fit} shows the best-fit NFW parameters for the NW and the SE mass structures. 
The main cluster is associated with the SE peak as shown in the mass map (figure \ref{fig:opt-gal-WL}).
The best-fit center positions of the SE and NW peaks agree with the BCG and the NW bright galaxy, SDSS J085020.60+001733.8, respectively. The mass ratio of the sub- to main- cluster is $0.32_{-0.23}^{+0.75}$, suggesting that the cluster is a major merger. The NW subcluster is offset by $\sim 700$ kpc from the main cluster center.

\begin{table}[htbp]
    \centering
    \tbl{Best-fit parameters obtained by the two-dimensional weak-lensing analysis.}{
    \begin{tabular}{ccc}\\\hline
Parameter&SE peak&NW peak\\\hline
\rule[3pt]{0pt}{8pt}RA [deg] &$132.635^{+0.005}_{-0.013}$&$132.588^{+0.006}_{-0.009}$\\
\rule[3pt]{0pt}{8pt}DEC [deg]&$0.256^{+0.005}_{-0.008}$&$0.292^{+0.012}_{-0.008}$\\
\rule[3pt]{0pt}{8pt}$M_{500}\,[10^{14}h_{70}^{-1}M_\odot]$  &$1.35_{-0.62}^{+0.76}$ & $0.52_{-0.36}^{+0.69}$ \\
\rule[3pt]{0pt}{8pt}$c_{500}$  &$2.56^{+0.17}_{-0.07}$&$2.80^{+0.51}_{-0.19}$\\
\rule[3pt]{0pt}{8pt}$r_{500}$ [arcmin/kpc]&$3.67^{+0.59}_{-0.68}/734^{+118}_{-136}$&$2.66\pm0.87/533\pm173$\\
\rule[3pt]{0pt}{8pt}$r_s$ [arcmin/kpc]&$1.43^{+0.28}_{-0.33}/287^{+56}_{-66}$&$0.95^{+0.40}_{-0.41}/190^{+80}_{-81}$\\\hline
\end{tabular}}
    \label{tab:WL_fit}
\end{table}


\subsection{X-ray data reduction}

Table \ref{tab:reduction} shows the information of the XMM-Newton archival data used in this paper.
The observation was carried out by the medium filter for MOS \citep{MOS} and the thin filter for pn \citep{pn}.

We reduced the data using the Science Analysis System (SAS) version 18.0.0, following the SAS threads\footnote{https://www.cosmos.esa.int/web/xmm-newton/sas-thread-esasimage}, and also Miyaoka et al. (2018). 
We filtered the raw data and extracted the cleaned events by the conditions of FLAG = 0 and $\rm PATTERN \leq 12$ for MOS, and FLAG = 0 and $\rm PATTERN \leq 4$ for pn.
The net clean exposure times became 50.9, 53.1, and 38.9 ks for the MOS1, MOS2, and pn data, respectively, from the raw exposure time (62.9~ks).
Then, point sources were removed using the SAS task \texttt{cheese}.
We also compared X-ray images with an optical image and manually removed the remaining point-like sources.
We used \texttt{mos\_spectra} and \texttt{pn\_spectra} to generate spectral files and redistribution matrix files (RMFs). 
As for the telescope responses, extended auxiliary response files (ARFs) were generated by using the \texttt{arfgen} command. 
Among the non-X-ray background (NXB), spectra and images of the quiescent particle background (QPB) were estimated and generated by using the \texttt{mos\_back} and \texttt{pn\_back} commands.
On the other hand, images of the soft proton contamination were generated by using \texttt{proton} command with the best-fit indices and normalizations, determined by spectral fitting using a broken power-law model for this background component.

\begin{table*}[htbp]
\centering\small
\tbl{XMM-Newton archival data. The flare time is removed from the exposure time.}{
\begin{tabular}{cccccc}
\hline
\hline
Obs ID&Target name&Start time [UT] & Exposure time (MOS1 / MOS2 / pn) [ks] & RA & Dec \\ \hline
761730501 & J0850+0022 &2015-10-18 09:45:40 & 50.9 / 53.1 / 38.9 & \timeform{08h50m27.88s} & \timeform{-00D22'55.0"}\\
\hline
\end{tabular}}
\label{tab:reduction}
\end{table*}

\subsection{X-ray image}

\begin{figure*}
    \begin{center}
    \includegraphics[width=8.0cm]{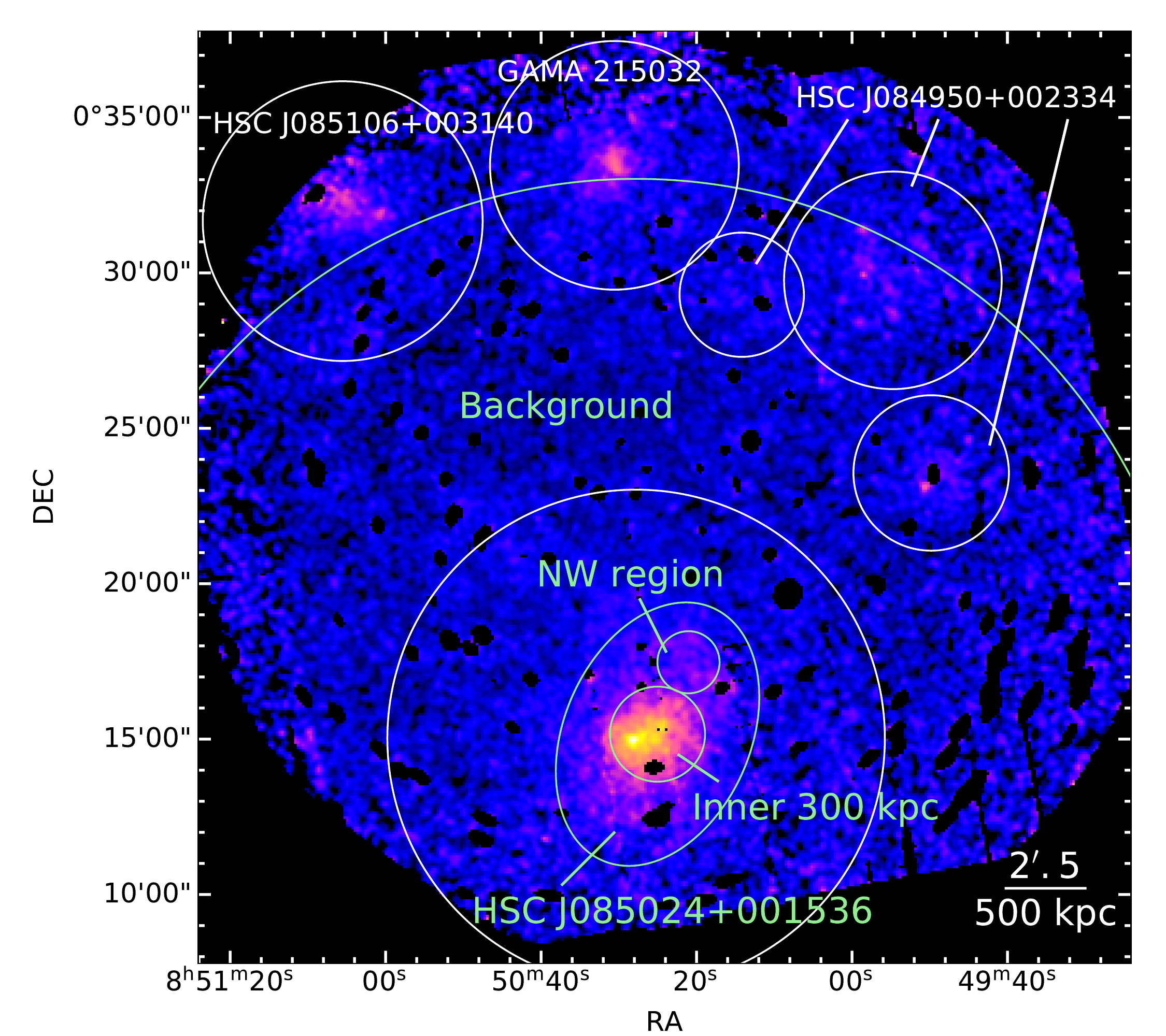}
    \includegraphics[width=8.0cm]{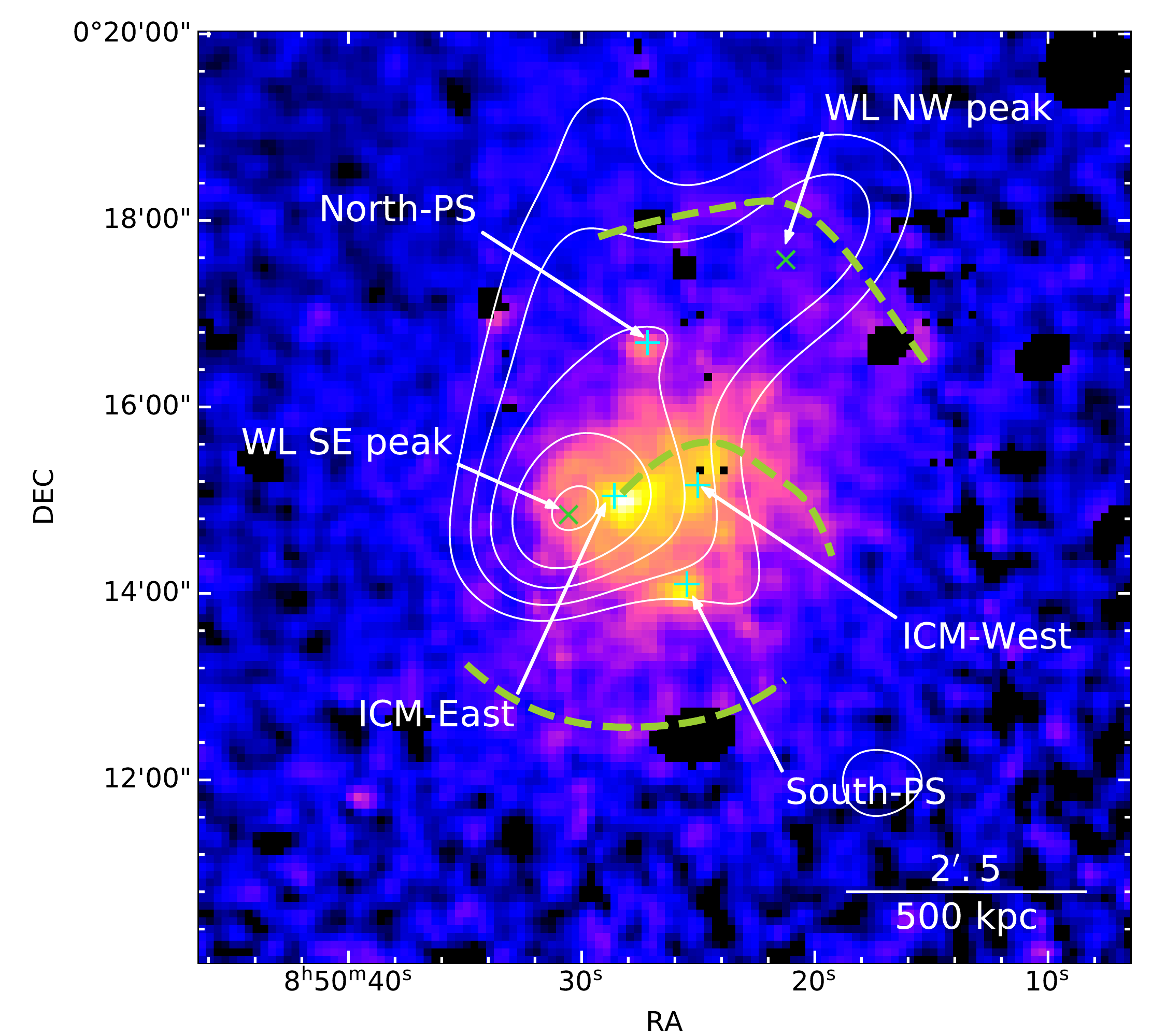}
    \end{center}
    \caption{(Left) X-ray image in the 0.4--4.0 keV band after subtracting the non-X-ray background (NXB), correcting exposure-time, and smoothing with $\sigma=5''$. MOS1, 2, and pn images were merged.
    The NXB includes quiescent particle background (QPB) and residual soft proton contamination, generated by the SAS tasks \texttt{mos\_back}/\texttt{pn\_back} and \texttt{proton}, respectively. HSC J085024+001536 is detected south in the field of view (green ellipse).
    HSC J084950+002334, HSC J085106+003140, and GAMA 215032 are also detected.
    Since HSC J084950+002334 is a low-redshift ($0.17$) and low-richness ($\sim15$) cluster \citep{Oguri18}, the member galaxies are widely and shallowly distributed. The three X-ray emissions are associated with the member galaxy distribution or the brightest cluster galaxy. As described in section 2.6, a region for sky background is shown with a green circle arc, excluding the regions shown with white circles.
    (Right) Close-up view of HSC J085024+001536. The weak-lensing mass contours are also shown
    with the peak positions as the green x marks (see figure~\ref{fig:opt-gal-WL} and
    table~\ref{tab:WL_fit}). Green crosses show positions of the leak-lensing (WL) mass map peaks, blue crosses show positions of the X-ray brightness peaks,
    and green dashed curves indicate the residual arc-like ridges. See section~\ref{subsec:ximg} for details.}
    \label{fig:X-gal_img}
\end{figure*}

The left panel of Figure~\ref{fig:X-gal_img} shows an XMM-Newton image in the 0.4--4.0~keV band.
Extended X-ray emission is clearly detected from HSC J085024+001536 for the first time.
The net counts after subtracting the QPB, in the elliptical region of HSC J085024+001536 defined in the left panel of figure~\ref{fig:X-gal_img}, are 5300 in the 0.3-11 keV band for the MOS2, and 12260 in the 0.4-11 keV band for the pn, respectively.
Note that three other clusters (HSC J084950+002334, HSC J085106+003140, and GAMA 215032) 
are also detected in the same field of view. 
The right panel of figure~\ref{fig:X-gal_img} shows a close-up view of HSC J085024+001536, 
overlaid with contours of the weak-lensing mass map shown in figure~\ref{fig:opt-gal-WL}.
The diffuse X-ray emission concentrates on the main cluster of the SE peak, and it is very faint around the NW peak of the weak-lensing mass map.

\subsection{Residual structures in the X-ray image}
\label{subsec:ximg}


We fit the X-ray surface brightness image in the 0.4--4.0 keV band to find evidence of a cluster merger.
We adopt a multi-component model to describe the complexity of X-ray surface brightness distribution. 
We consider two components of diffuse emission (hereafter, ICM-West and ICM-East)\footnote{We also fitted the image with a single $\beta$ model and three $\beta$ models (plus point sources and a constant background model as shown in equation~(\ref{eq:SX}). In the one $\beta$ case, the C-statistic value was 64512 and large residuals still remained near the center, compared with 64339 for the two $\beta$ case. In the three $\beta$ case, the C-statistic value, 64309, was comparable to that for the two $\beta$ case, and the global structures were the same. Therefore, the two $\beta$ case is valid.}
Each component is described with a two-dimensional elliptical $\beta$ model for diffuse emission,
\begin{eqnarray}
  S_{\beta}^e(x,y)&=&S_0\left(1+\frac{X(x,y)^2(1-e)^2+Y(x,y)^2}{r_c^2(1-e)^2}\right)^{-3\beta+1/2}\\
  X(x,y)&=&(x-x_c)\cos\theta+(y-y_c)\sin\theta\\
  Y(x,y)&=&(y-y_c)\cos\theta-(x-x_c)\sin\theta
\end{eqnarray}
where $r_c$ is the core radius, $\beta$ is the outer slope, $e$ is the ellipticity, $\theta$ is the orientation angle and $x_c,y_c$ are the coordinates of the center position. The parameter range for the center position is limited within
$50''\times 50''$ square boxes from the X-ray peaks. 
Since the outer slope $\beta$ and the ellipticity $e$ for the ICM-East component cannot be constrained, we fix $\beta=1$ and $e=0$, respectively. We additionally introduce delta functions ($S_{\rm PS}$) to represent two point sources 
(hereafter, North-PS and South-PS), and a constant ($S_b$) to represent the sky-background component.
The total surface brightness distribution, $S_X$, is described by 
\begin{eqnarray}
  S_X=S_{\beta,{\rm West}}^e+S_{\beta,{\rm East}}^e+S_{{\rm PS, North-PS}}+S_{\rm PS, South-PS} + S_b.
  \label{eq:SX}
\end{eqnarray}
We use Sherpa \citep{Sherpa} in CIAO version 4.11 \citep{CIAO} for image fitting. 
We determine a fitting area to 12$'$ square region centered at the BCG.
Source distribution models were convolved with an exposure map, and a telescope's point spread function (PSF) map which was generated at the energy of 2.2 keV and at the BCG position 
using the \verb+psfgen+ tool. 
The observed images, PSF maps, exposure maps, and NXB images were rebinned so that the pixel size becomes $5''$.
The images of the three detectors were fitted simultaneously using the C-statistic \citep{Cash}.
The normalization of each component and the constant for the background were linked between MOS1 and MOS2 images because the cluster center is on the dead chip of MOS1.

\begin{figure}
    \begin{center}
    \includegraphics[width=8.0cm]{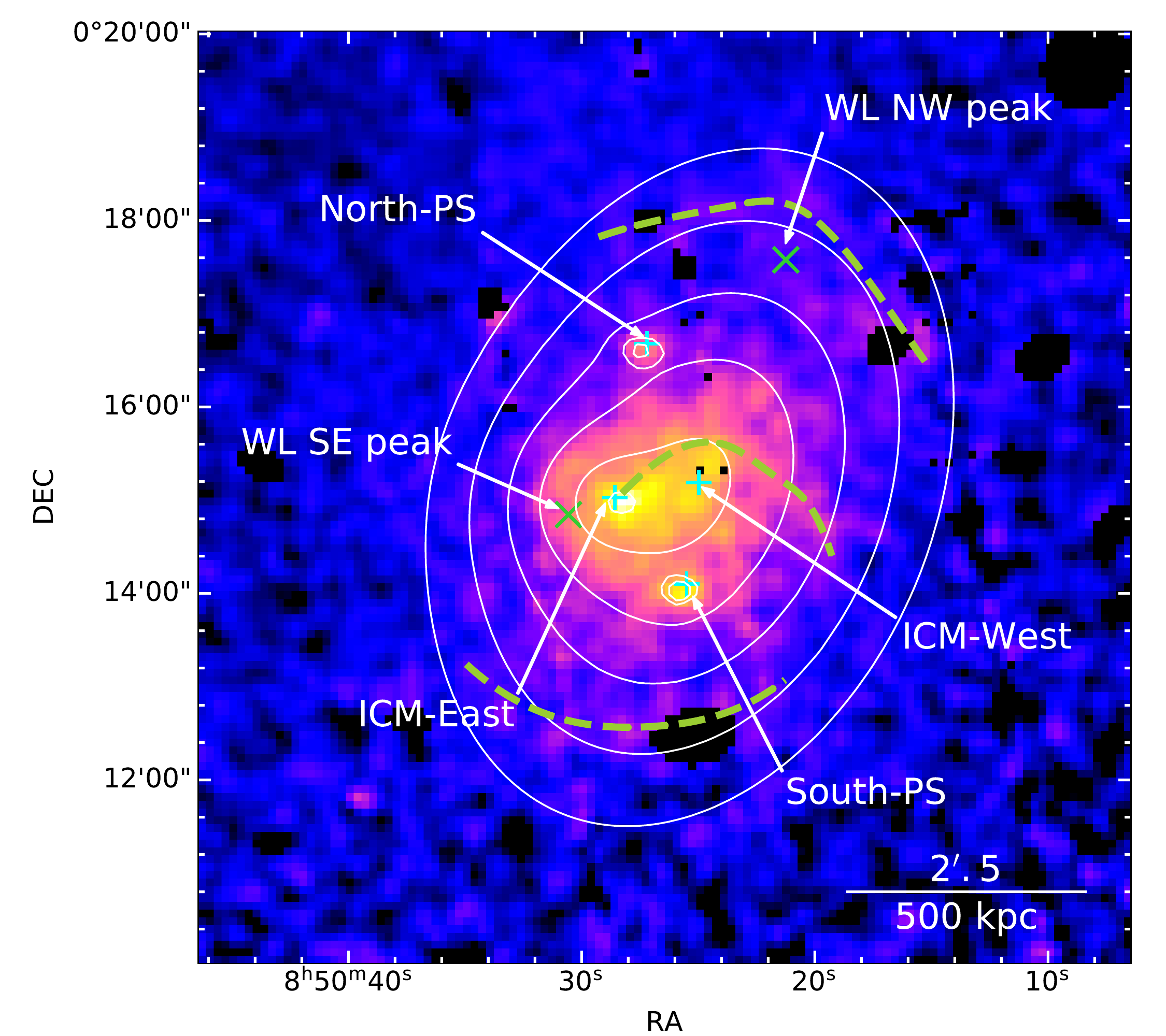}
    \end{center}
    \caption{X-ray image in the $0.4-4.0$~keV band, overlaid with contours of the best fit models described in table~\ref{tab:image_fit}, convolved with the telescope PSF, and smoothed with $\sigma=5"$. The X-ray image is the same as the right panel of figure~\ref{fig:X-gal_img}.
    }
    \label{fig:best_img}
\end{figure}

Table \ref{tab:image_fit} summarizes the best-fit parameters of the image fitting, and figure~\ref{fig:best_img} shows contours of the best-fit models overlaid with the X-ray image.
The best-fit positions of ICM-West, ICM-East, North-PS, and South-PS are also shown in figure \ref{fig:best_img}. The ICM-East coincides with the BCG position, and the ICM-West is the X-ray morphological center.
Since instrumental Al K$\alpha$ and Si K$\alpha$ lines are predominant in the low surface brightness region, we evaluated the influence of these lines by excluding the 1.3--1.9 keV energy range. As a result, the C-statistic was greatly improved (C-statistic$=64339\rightarrow62146$, d.o.f unchanged), but the best-fit values were unchanged within errors.
When we fitted X-ray images 
after subtracting images of the soft proton contamination generated by the SAS task \texttt{proton}, the parameters of the $\beta$ models were unchanged
within errors because the observational data is not significantly affected by the flare.
Note that we confirmed that the position and energy dependence of the PSF shape has a negligible influence on the fitting results.

\begin{table*}[htbp]
    \centering
    \tbl{Best-fit parameters obtained by fitting the X-ray image.}{
    \begin{tabular}{cccccccc}\\\hline
Parameter&ICM-West&ICM-East&North-PS&South-PS&Background\\\hline
\rule[3pt]{0pt}{8pt}$r_c$~[arcsec~/~kpc]&$68^{+10}_{-5}$ / $223^{+34}_{-15}$&$43\pm 4$ / $140^{+12}_{-13}$&-&-&-\\
\rule[3pt]{0pt}{8pt}$\beta$&$0.51^{+0.04}_{-0.02}$&1.00(fix)&-&-&-\\
\rule[3pt]{0pt}{8pt}$e$&$0.32\pm 0.02$&0.00(fix)&-&-&-\\
\rule[3pt]{0pt}{8pt}$\theta$~[rad]&$1.16\pm 0.04$&0.00(fix)&-&-&-\\
\rule[3pt]{0pt}{8pt}RA~[deg]&$132.6043\pm0.0001$&$132.6191\pm0.0001$&$132.6132^{+0.0011}_{-0.0003}$&$132.6062^{+0.0012}_{-0.0002}$&-\\
\rule[3pt]{0pt}{8pt}Dec~[deg]&$0.2527\pm0.0005$&$0.2507\pm0.0005$&$0.2781^{+0.0002}_{-0.0012}$&$0.2350^{+0.0002}_{-0.0012}$&-\\
\rule[3pt]{0pt}{8pt}Norm$_{\rm MOS1}$~[c\,s$^{-1}$\,deg$^{-2}$]&$14.1^{+0.9}_{-1.8}$&$15.8^{+2.3}_{-1.9}$&$159^{+43}_{-39}$&$220^{+56}_{-53}$&\multicolumn{1}{c}{$0.73^{+0.04}_{-0.05}$}\\
\rule[3pt]{0pt}{8pt}Norm$_{\rm MOS2}$~[c\,s$^{-1}$\,deg$^{-2}$]&$13.9^{+0.8}_{-1.6}$&$\dagger$&$\dagger$&$\dagger$&\multicolumn{1}{c}{$\dagger$}\\
\rule[3pt]{0pt}{8pt}Norm$_{\rm pn}$~[c\,s$^{-1}$\,deg$^{-2}$]&$11.8^{+0.6}_{-1.1}$&$11.7^{+1.2}_{-1.0}$&$96^{+19}_{-18}$&$197^{+31}_{-29}$&\multicolumn{1}{c}{$0.83\pm 0.03$}\\
\hline
C-statistic&\multicolumn{5}{c}{64339.3}\\
d.o.f&\multicolumn{5}{c}{41669}\\
C-statistic/d.o.f&\multicolumn{5}{c}{1.544}\\\hline
\multicolumn{6}{l}{$\dagger$ Norm (MOS2, W535-West) $\times$ Norm (MOS1, each component) $/$ Norm (MOS1, W535-West)}\\
\end{tabular}
}
    \label{tab:image_fit}
\end{table*}

\begin{figure*}
    \begin{center}
    \begin{minipage}{0.49\hsize}
    \includegraphics[width=8.0cm]{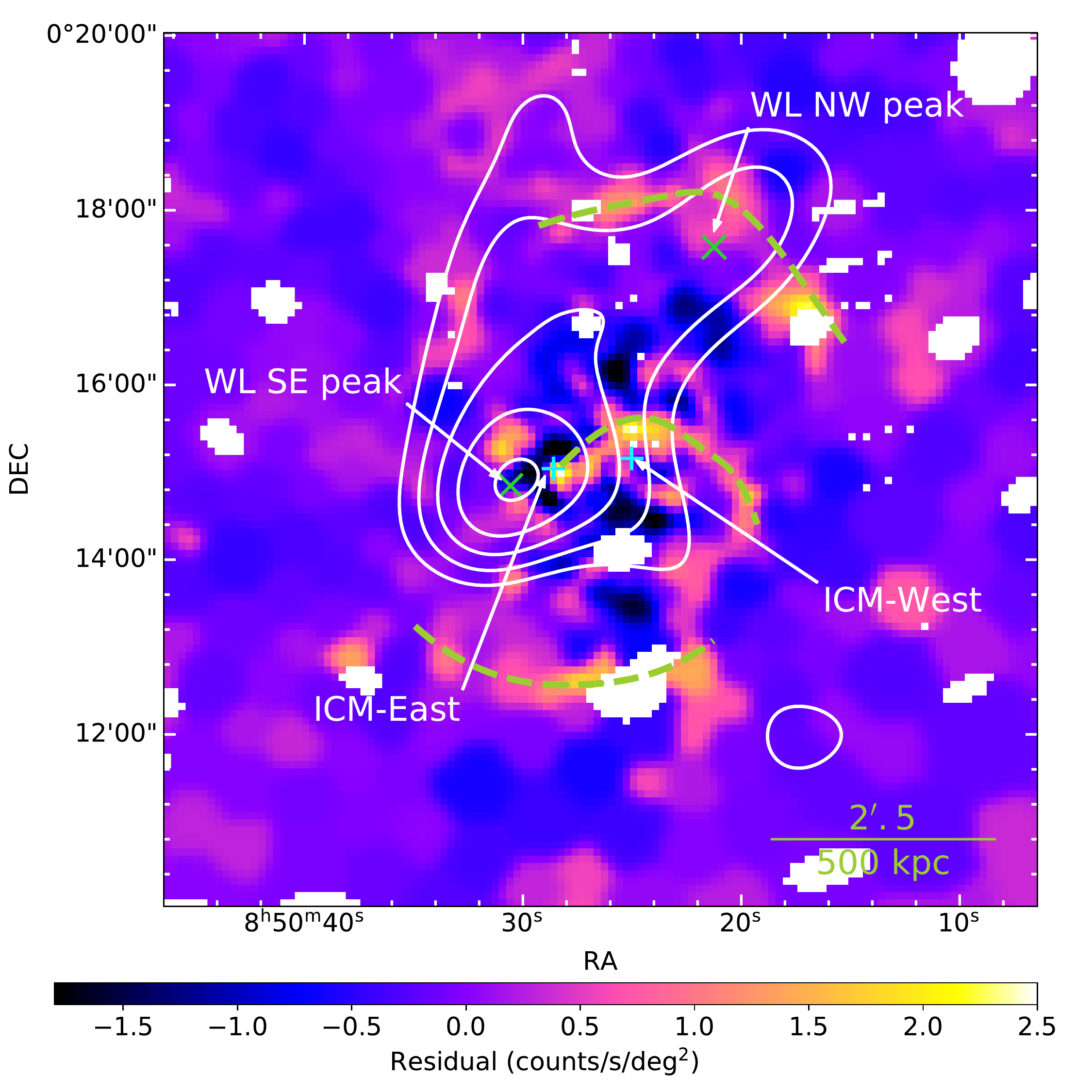}
    \end{minipage}
    \begin{minipage}{0.5\hsize}
    \includegraphics[width=8.0cm]{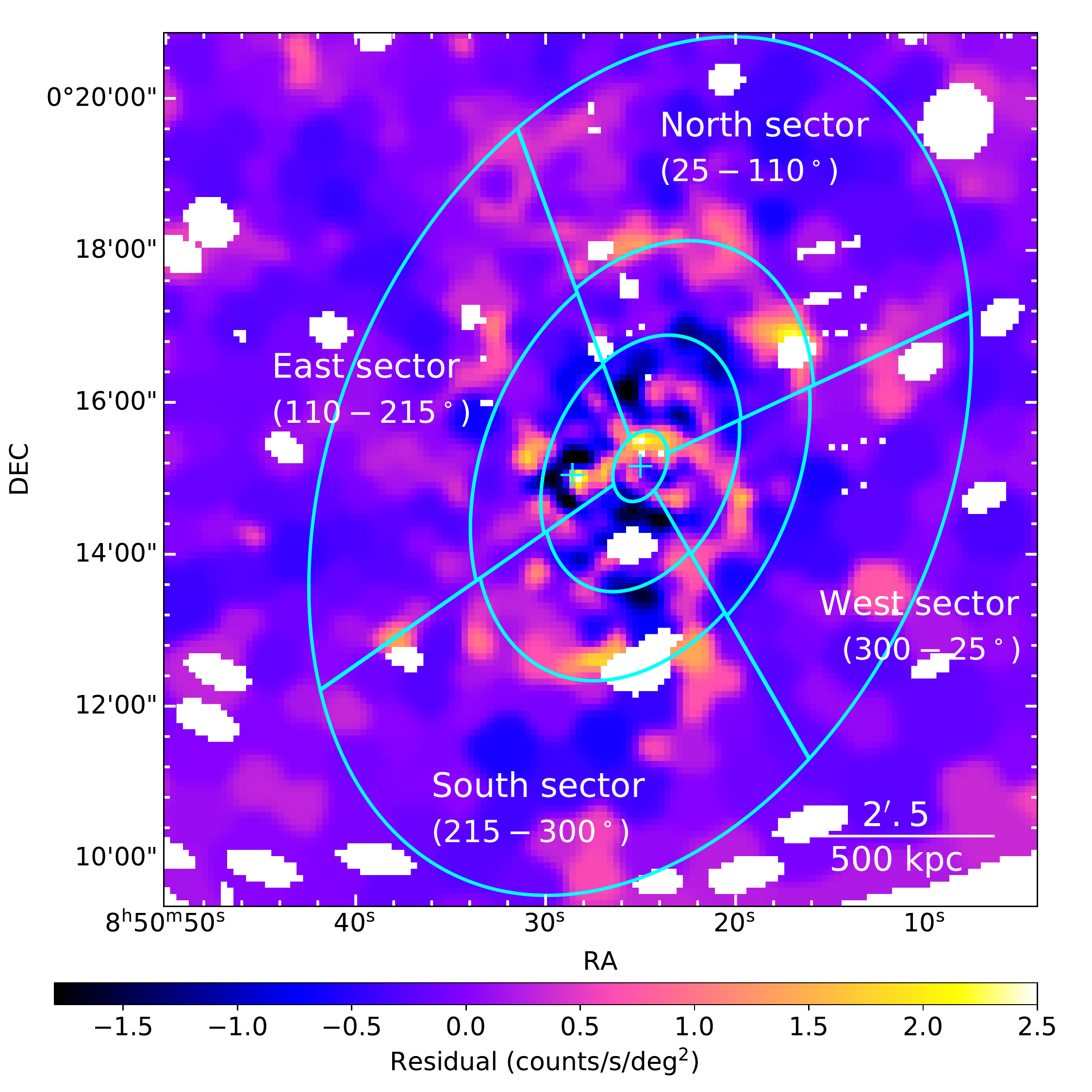}
    \end{minipage}
    \end{center}
    \caption{(Left) Residual map after subtracting the best-fit models shown in table~\ref{tab:image_fit} and figure~\ref{fig:best_img}
    from the X-ray image.
    We stacked MOS1, MOS2, and pn residual images. The stacked residual image is binned into a Weighted Voronoi Tesselation (WVT) algorithm \citep{Diehl2006}, where each region contains at least 40 counts in the original image. 
    The stacked residual image is smoothed with $\sigma=5''$ for visual purpose. 
    White contours show the weak-lensing mass map, which are the same as those in the right panel of figure~\ref{fig:X-gal_img}. 
    Light blue crosses show the best-fit positions of ICM-West and ICM-East. 
    Green dashed curves are added to emphasize the residual arc-like ridges. 
   (Right) Definition of regions. Four sectors are defined taking into account locations of the residual arc-like ridges. They are referred to as north ($25-110^\circ$), east ($110-215^\circ$), south ($215-300^\circ$), and west ($300-25^\circ$). Each sector is further divided into four annular regions in spectral analysis in section \ref{sec:radial_profile}. 
    }
    \label{fig:img_residual}
\end{figure*}

The left panel of figure \ref{fig:img_residual} shows the residual image after subtracting the best-fit model shown in table \ref{tab:image_fit} and in figure~\ref{fig:best_img} from the X-ray image.
There are arc-like residual structures in the cluster center, and to the north and to the south of the cluster center\footnote{To check the robustness of the detected excesses with respect to the underlying model, we tested $\beta$ of the ICM-East fixed at 0.7, and $\beta$ free. We also tested that the ellipticity $e$ set free, but $e$ became 0. In all cases, the arc-like structures were unchanged. They were also unchanged in the case of a single $\beta$ model and three $\beta$ models. In fact, these features are weakly visible in the raw image (see the right panel of figure~\ref{fig:X-gal_img}).}.
Note that these structures are clearly found in the residual image of each detector. We confirmed that the significance levels of these structures are $\sim1\sigma$ for the central structure, and $\sim2-3\sigma$ for the north and the south structures in the stacked residual image, respectively.
Based on the location of the residual arc-like structures, four sectors were defined as shown in the right panel of figure \ref{fig:img_residual}; north ($25-110^\circ$), east ($110-215^\circ$), south ($215-300^\circ$), and west ($300-25^\circ$).

\begin{figure*}
    \begin{center}
    \begin{minipage}{0.49\hsize}
    \includegraphics[width=8.0cm]{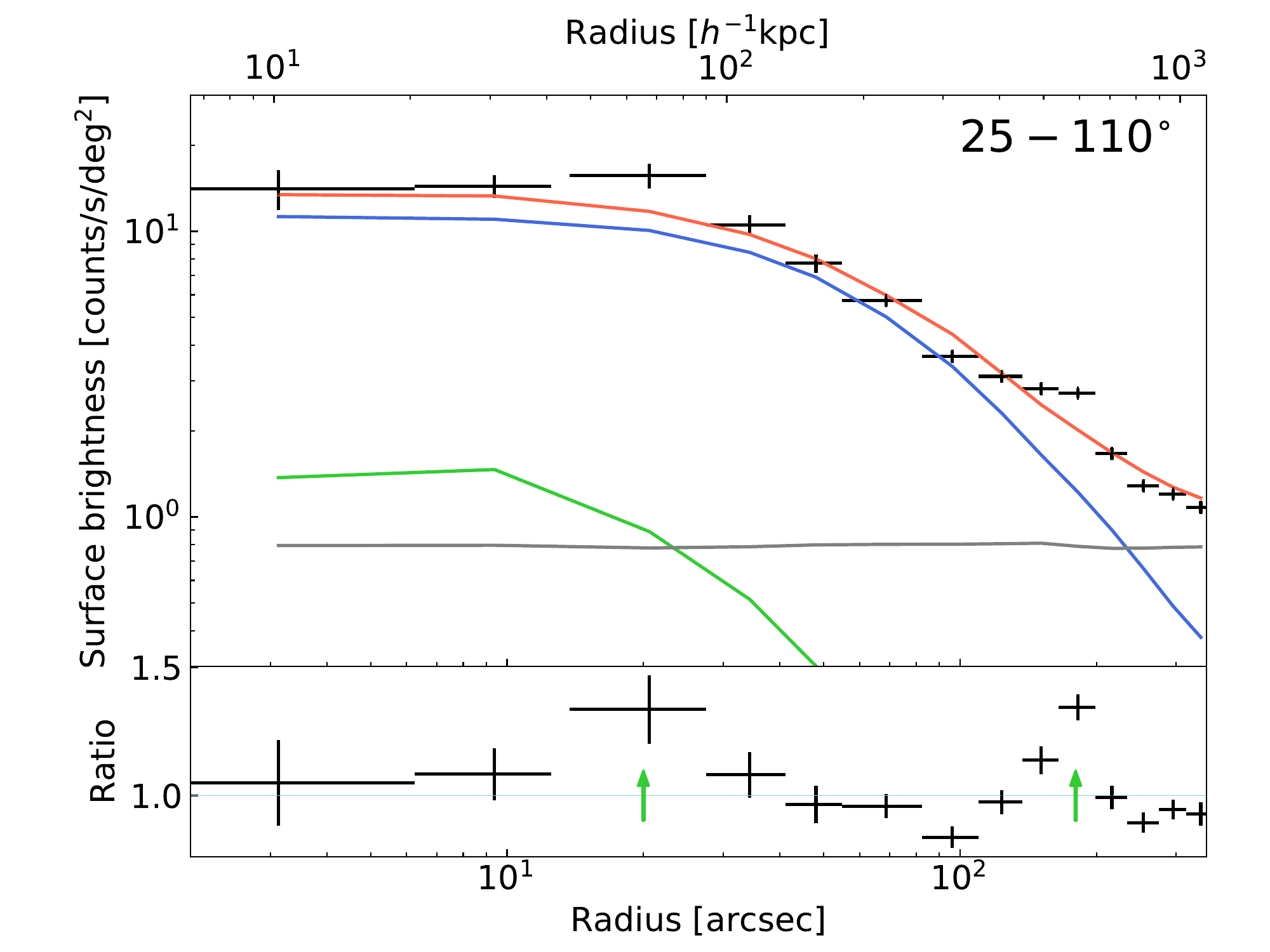}
    \end{minipage}
    \begin{minipage}{0.5\hsize}
    \includegraphics[width=8.0cm]{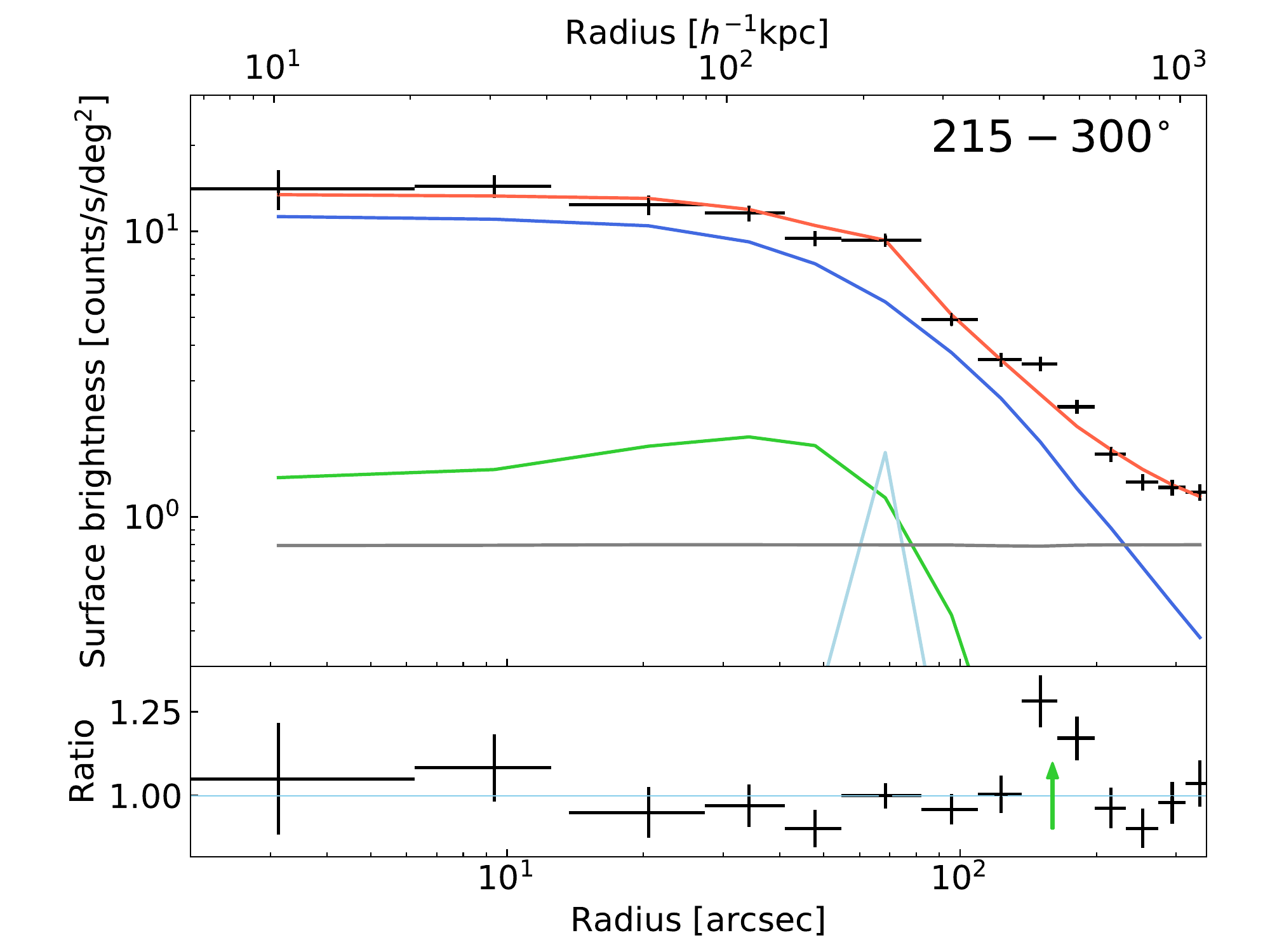}
    \end{minipage}
    \end{center}
    \caption{The X-ray surface brightness profiles in the north sector (left) and in the south sector (right) defined in the right panel of figure~\ref{fig:img_residual}, as a function of projected distance from the ICM-West center.
    The NXB was subtracted, the exposure time was corrected, and MOS1, 2, and pn data were merged.
    The blue, green, light blue, and gray solid curves show the contribution of ICM-West, ICM-East, South-PS, and sky-background components, respectively, which are calculated using the best-fit models shown in table~\ref{tab:image_fit} and figure~\ref{fig:best_img}. The red curves are their sum. The lower panel of each figure shows the residuals between the data and the red model curve. The green arrows show two excesses in the north sector and one excess in the south sector. 
    }
    \label{fig:SB_radial-profile}
\end{figure*}

Figure~\ref{fig:SB_radial-profile} shows radial profiles of the X-ray surface brightness after subtracting 
the NXB, in the north and the south sectors shown in figure~\ref{fig:img_residual}. 
Large excess is shown at the distance of 40--90~kpc and 450--650~kpc in the north sector, and at 450--650~kpc in the south sector. 
The significance levels of these excesses are $2.5\sigma$, $6.6\sigma$, and $3.6\sigma$, respectively.
The excess ridges are located at $\sim80$ kpc and $\sim 550$ kpc from the ICM-West (figure~\ref{fig:img_residual}). The south and the north ridges are tangentially elongated along $\sim 400$ and $\sim 650$ kpc length, respectively.
Although these excesses can be represented by the cold front density jump model given by equation (A4) of \citet{Owers09}, the parameters were not well constrained due to limited photon statistics.

\subsection{Spectral features of the ICM}
\label{sec:spec}


To understand the ICM properties, we first fit the EPIC spectra of the three regions shown in the left panel of figure \ref{fig:X-gal_img}.
The first is an ellipsoid region centered at the ICM-West, with the semi-major axis of 4.4$'$, the 
semi-minor axis of 3$'$, and tilted by 1.16~rad ($66^\circ$) based on the X-ray image fitting result shown in table~\ref{tab:image_fit}.
We refer to it as the entire region of the cluster. The second is a circle of 300~kpc radius centered at the ICM-West center, referred to as the inner 300 kpc.
The third is a circular NW region of $1'$ radius from the best-fit NW mass peak position determined by the weak-lensing analysis, referred to as the NW region. 
The background region is defined as an annular region with a radius of 8$'$--18$'$ from the BCG,
but excluding regions containing other objects as shown with white circles in figure~\ref{fig:X-gal_img}.

Among the NXB, the quiescent particle background (QPB) spectra were estimated using the SAS tasks \verb+mos+\_\verb+back+ and \verb+pn+\_\verb+back+. 
The energy band is set to be 0.3--11.0 keV and 0.4--11.0 keV for the MOS and pn spectra, respectively.
ARFs for extended sources were made, assuming that the surface brightness is constant within each region.
Spectral files are rebinned so that each bin contains at least 10 counts for the entire cluster region,
1 count for the NW region, and 20 counts for the background region, respectively.

XSPEC version 12.11.0 \citep{XSPEC} was used for spectral fitting, with C-statistic for finding the best-fit parameters. We adopt the APEC version 3.0.9 \citep{APEC} as a model for emission from the ICM, 
with solar abundance table by \citet{lodders2009}. As a photoelectric absorption model, \verb+phabs+ is utilized
with the column density fixed at $3.19\times10^{20}~{\rm cm^{-2}}$ based on the LAB survey \citep{LABsurvey}.
Following \citet{Snowden}, background components of the Local Hot Bubble (LHB), the Milky Way Halo (MWH), the Cosmic X-ray Background (CXB), and NXB contribution other than the QPB, i.e., instrumental lines of Al K$\alpha$, Si K$\alpha$ for MOS, and Al K$\alpha$, Cu K$\alpha$, K$\beta$, Ni K$\alpha$, Zn K$\alpha$ for pn, and residual soft proton contamination
\citep{MOS_background1,MOS_background2,pn_background,XMM_SWCX1,XMM_SWCX2}, are represented by models.
Spectra of the source regions and the background regions are fitted simultaneously.
Parameters among detectors were linked.
The redshift was fixed at the cluster photometric redshift $z=0.1966$ \citep{Oguri18}. 

\begin{figure}
    \begin{center}
    \includegraphics[width=8.0cm]{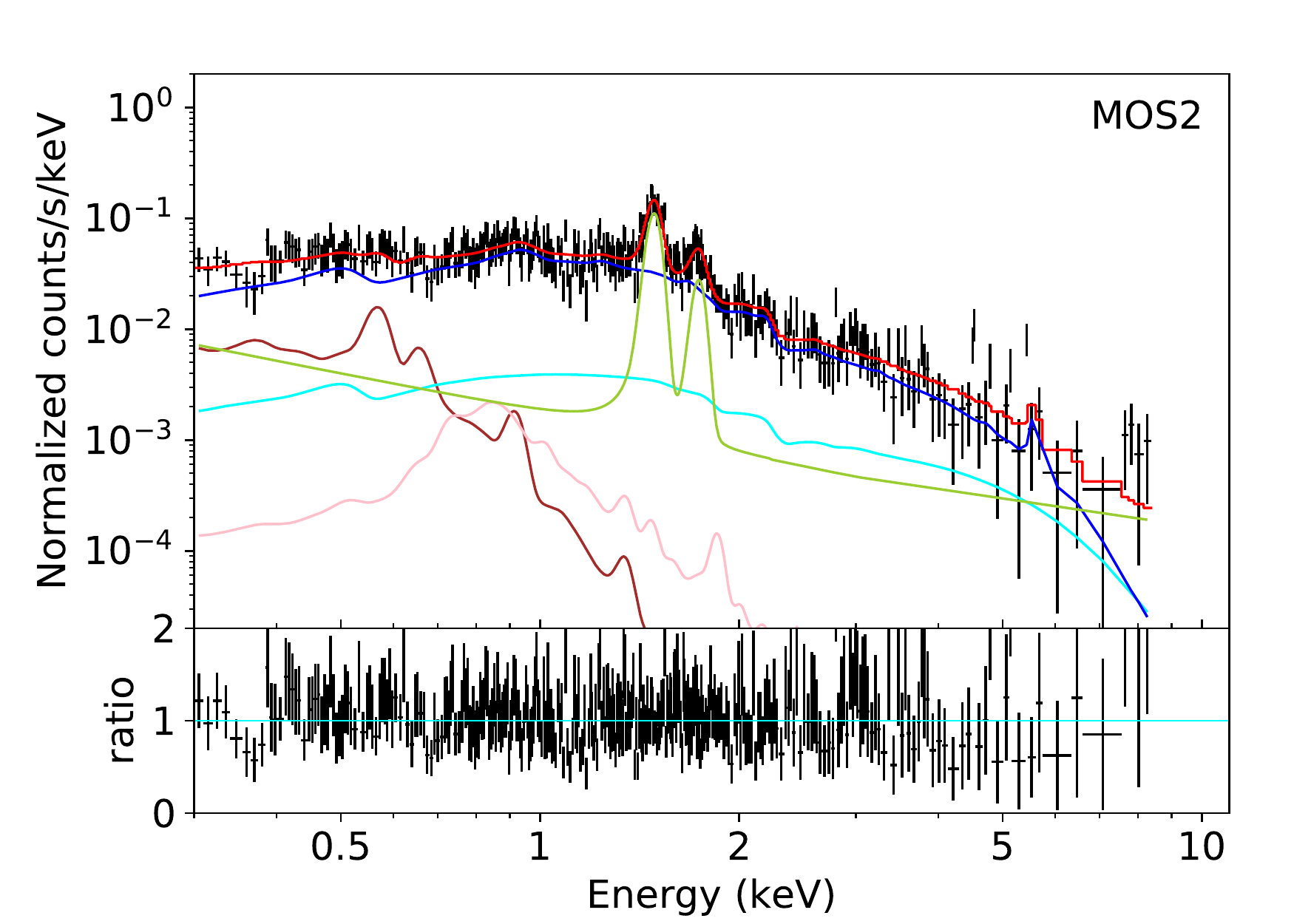}
    \end{center}
    \caption{QPB-subtracted spectra and best-fit models of the MOS2 from the entire cluster region. The blue, brown, pink, light-blue, and yellow-green spectra show the ICM, Local Hot Bubble (LHB), Milky Way halo (MWH), cosmic X-ray background (CXB), and unsubtracted NXB (instrumental fluorescent lines and residual soft proton contamination) component, respectively. The red line shows the sum of them.}
    \label{fig:spectra}
\end{figure}

Figure \ref{fig:spectra} shows the spectra and the best-fit models of the entire cluster region, and
table \ref{tab:spec_fit} summarizes the best-fit parameters of the ICM component in the entire, the inner 300~kpc and the NW regions.
The best-fit temperature of the entire ICM was $3.15\pm0.14$~keV. 
Note that, when we adopted a two-temperature model,
we could not determine the temperature ($kT_{1}=1.70^{+0.37}_{-0.39}$~keV, $kT_{2}<4.27$~keV),
and the C-statistic was almost unchanged from the 1$T$ model fitting
(1781.45$\rightarrow$1778.52). 
The temperature of the NW region is $kT=2.06^{+0.35}_{-0.26}$~keV, which is lower than that of the entire ICM. 
Note that this region is also contaminated with a component of the residual arc-like structure described in
section~\ref{subsec:ximg}, and its contribution cannot be ignored. Thus, we cannot clearly constrain
the spectral parameters of the ICM associated with the NW peak.

\begin{table}[htbp]
    \centering
    \tbl{Best-fit spectral parameters of the entire, inner 300 kpc and the NW region.}{

\begin{tabular}{cccc}\\\hline
Parameter&Entire&Inner 300 kpc&NW region\\\hline
\rule[3pt]{0pt}{8pt}kT [keV]&$3.15\pm 0.14$&$3.42^{+0.25}_{-0.17}$&$2.06^{+0.35}_{-0.26}$\\
\rule[3pt]{0pt}{8pt}Abundance [Z$_\odot$]&$0.32\pm0.074$&$0.36^{+0.10}_{-0.09}$&$0.58^{+0.33}_{-0.23}$\\
\rule[3pt]{0pt}{8pt}Redshift&\multicolumn{3}{c}{0.1966 (fixed)}\\
\rule[3pt]{0pt}{8pt}norm$(10^{-6})$&$8.56^{+0.32}_{-0.31}$&$44.8\pm2.0$&$17.0^{+3.0}_{-2.8}$\\
\hline
\multicolumn{1}{c}{C-statistic}&1781.45&1941.51&1407.56\\
\multicolumn{1}{c}{d.o.f}&2046&2334&1593\\
\multicolumn{1}{c}{C-statistic / d.o.f}&0.871&0.832&0.884\\\hline
\end{tabular}
}
    \label{tab:spec_fit}
\end{table}

\subsection{Radial profiles of temperature, density, pressure and entropy}
\label{sec:radial_profile}

We carry out spectral fitting considering the projection effect, and compute three-dimensional radial profiles of the temperature, the electron density, the gas pressure, and the entropy in the four sectors defined in figure \ref{fig:img_residual}. 
Each sector is divided into four annular regions.
We adopt a full ellipse for the innermost region because of limited number of photons and use in total 13 regions for spectral fitting.
For each projected region, we calculate volumes of the shells in the same line of sight to the region. 
In deprojection technique, we assume the following geometry; 
a prolate spheroid aligned with the major and minor axes of the best-fit ellipse (Table \ref{tab:image_fit}) on the sky plane and the uniform ICM distribution in each shell of the spheroid. 
The dead area and the bad pixels of the detectors are properly excluded from the volumes. Weighting factors are calculated by volume ratios, and the 13 spectral files are fitted simultaneously by multiplying  the weighting factors.
The abundance is linked between all regions.

\begin{figure*}
    \begin{center}
    \includegraphics[width=16.0cm]{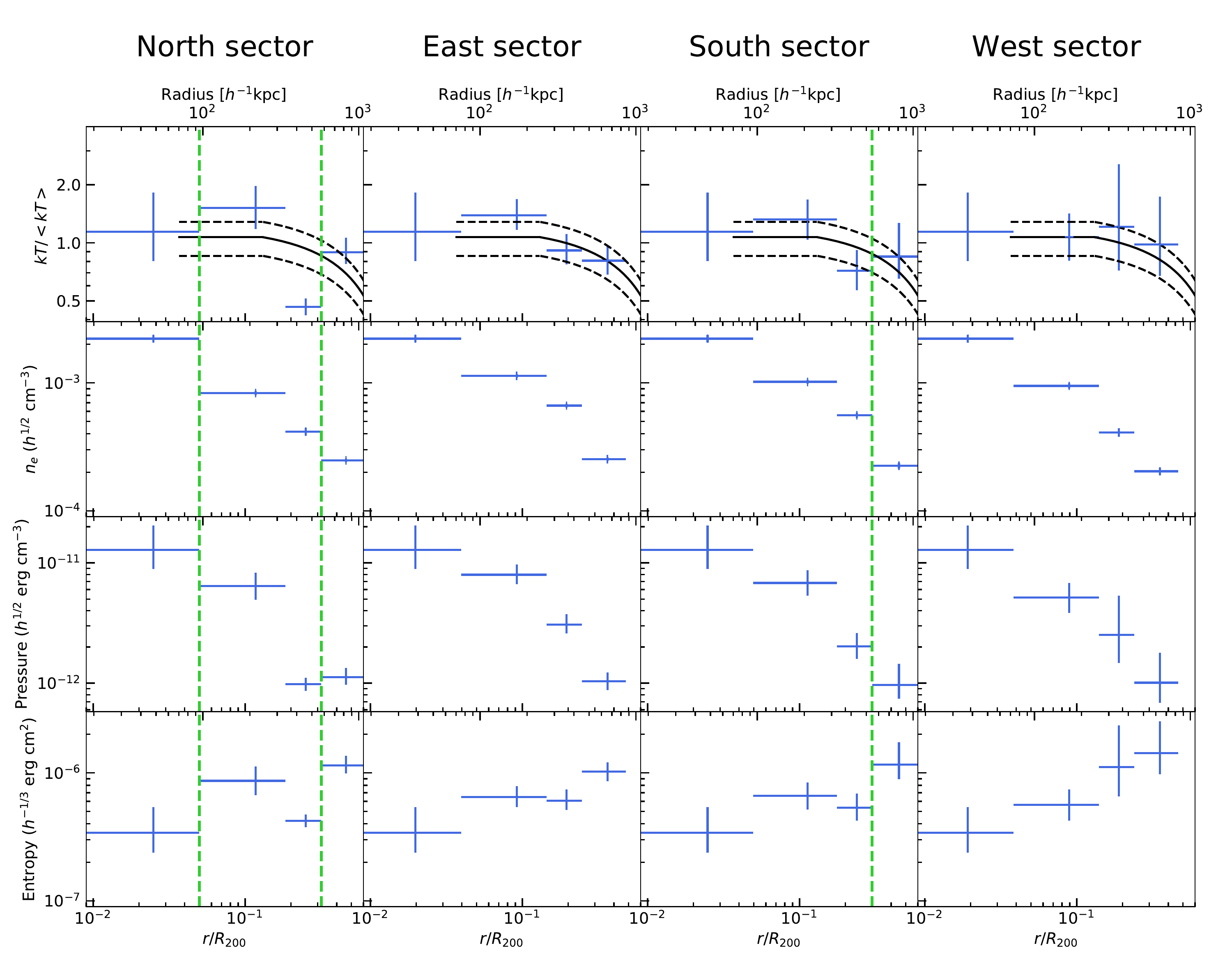}
    \end{center}
       \caption{Three-dimensional radial profiles of the temperature scaled with the average ICM temperature $<kT>=3.15$ keV (top), 
       electron density (middle-top), gas pressure (middle-bottom), and entropy (bottom). 
       From left to right, the panels show the north, east, south and west sectors, respectively.  
       The radii are normalized by the overdesnity radus $r_{200}=1784$ kpc of the main cluster.
       In the top panels, the black sold curves are the normalized temperature profiles of nearby, relaxed clusters \citep{Vikhlinin2005} for a reference, and the dashed curves are its 20\% scatters.
       The dashed green lines are the positions of the residual arc-like structure of the surface brightness on the sky.
       }
    \label{fig:kT_profile}
\end{figure*}

Figure \ref{fig:kT_profile} shows the resulting three-dimensional profiles for gas properties in the four sectors. 
We compare the temperature profiles in the east sector (top second left panel) and the west sector (top right panel) with the averaged temperature profile of 13 nearby, relaxed clusters \citep[][; eq. 2]{Vikhlinin2005} with $<kT>=3.15$~keV and find an agreement.
In contrast, the temperature in the $350-550$~kpc region of the north sector ($1.47^{+0.15}_{-0.14}$~keV) is lower than 
the temperature in the 550-1000 kpc region of the same sector ($2.82^{+0.54}_{-0.37}$~keV), and than the averaged profile for nearby, relaxed clusters \citep{Vikhlinin2005}.
The low-temperature region corresponds to the region containing the inner half of the arc-like residual structure, i.e., the region inside the candidate front, shown in figures~\ref{fig:img_residual} and \ref{fig:SB_radial-profile}. The gas pressures 
($P_{\mathrm{in}}=9.82^{+1.25}_{-1.17}\times10^{-13}~\mathrm{erg~cm^{-3}}$
and $P_{\mathrm{out}}=1.12^{+0.22}_{-0.15}\times10^{-12}~\mathrm{erg~cm^{-3}}$)
are nearly constant across the outer residual ridge. On the other hand, the temperatures across the inner residual ridge of the north sector and across the outer residual ridge of the south sector are nearly constant. However, the entropy across all the ridges changes by a factor of two (bottom panels).

\section{Discussion}
\label{sec:discuss}

\label{subsec:ColdFront}

The residual map of the X-ray surface brightness (figure~\ref{fig:img_residual}) and complementary radial profiles (figure~\ref{fig:SB_radial-profile}) show three arc-like excess patterns. The two residual ridges are located at $\sim550$ kpc from the ICM-West center in the north and south regions, and the one ridge is close to the center at $\sim80$ kpc in the north region. The deprojected temperatures across the ridges (figure~\ref{fig:kT_profile}) present that the inner temperatures are likely to be somewhat lower than the outer ones. The largest temperature change occurs at the northern outer ridge, showing $T_{\mathrm{in}}/T_{\mathrm{out}}=0.52^{+0.09}_{-0.11}$. The changes in the other ridges are not significant. Despite the diversity in the temperature properties, the entropy across all ridges change by more than a factor of two. It indicates that different originated gases contact around the ridges. The entropy profile in the north direction increases out to $\sim300$ kpc, decreases at $\sim400$ kpc, and increases beyond $\sim550$ kpc. This feature is convectively unstable and thus expected to have a short lifetime. 
The disturbed gas feature would be dissipated at the order of the sound-crossing time. 
Given the spatial scale of the entropy pattern, $\sim550$ kpc, and the sound-velocity,
$c_s=616^{+32}_{-30}\,{\rm kms^{-1}}$, its age is expected to be $0.9\pm0.1$ Gyr or longer.  
A perturber inducing the disturbance in the surface brightness is likely to be the NW subhalo.

We estimate a gas mass fraction for the NW subhalo within a radius of 1$'$, which is comparable to $r_{2500}$, from the NW peak of the weak-lensing mass map. The weak-lensing mass for the NW subhalo is $M_{\mathrm{WL}}=1.7^{+1.4}_{-1.0}\times10^{13}~M_\odot$ from the two-dimensional weak-lensing analysis (\S\ref{sec:WL}). 
The gas density is estimated from the normalization of the APEC model shown in table~\ref{tab:spec_fit}, assuming that it is constant within the same sphere and the filling factor of 1. Then, the gas mass in the sphere is estimated to be $M_{\mathrm{gas}}=6.6^{+0.6}_{-0.5}\times10^{11}~M_\odot$.
The gas mass fraction in the NW subhalo, $f_{\mathrm{gas}}=4.0^{+2.3}_{-3.3}\%$, is marginally smaller 
than expected by numerical simulation with AGN feedback, $f_{\mathrm{gas}}=7.6^{+0.4}_{-0.5}\%$, computed by the weak-lensing mass and the redshift \citep{Planelles2013}. It indicates that a large fraction of the ICM, which had been associated with the NW subhalo, was stripped by the cluster collision. 
Note that the gas mass fraction obtained here should be regarded as an upper limit, since there is ICM contributions from the main cluster and the north ridge in this region.

The symmetric distribution of the arc-like features around the ICM-West center strongly suggests that these features are due to the sloshing of the central ICM of the main cluster, as suggested by \citet{2006ApJ...650..102A}. In this scenario, the NW subcluster could play the role of a gas-poor subcluster which triggered the disturbance of the gravitational potential including the gas motions.

We estimate a merger time scale from the weak-lensing masses using a simple 
gravitational interaction \citep{Ricker01}. 
We use $r_{200}$ and $M_{200}$ of the SE halo and the NW subhalo in the calculation. 
We assume a large impact parameter $b=500$~kpc, since there are features that resemble cold fronts in the simulations with a large impact parameter \citep[e.g.,][]{Ricker01,Poole06}.
Given these conditions and using equation (31) of \citet{Ricker01}, a relative velocity of the two subclusters is $\sim2000~\mathrm{km~s^{-1}}$ at the closest impact.
Note that the relative line-of-sight velocity of the two subclusters determined from the redshift of the member galaxies (\S\ref{subsec:galden}) is $\sim1300~\mathrm{km~s^{-1}}$,
and there is no major contradiction with our simple estimation. 
In this case, it takes $\sim0.12$~Gyr for the subhalo to reach the current distance ($\sim700$~kpc) after the closest impact. 
This estimate is much shorter than the age of the cold front $\gtrsim 0.9$ Gyr (see the first paragraph in this section). 
Since the dynamical time from the closest separation to the maximum expansion is $\sim 1.6$ Gyr, which is calculated equation (31) of \citet{Ricker01}, the total time after the first closest impact is $\simgt 3$ Gyr. Therefore, the cluster is likely to be in the late phase of the cluster merger, at least at the second impact. Again, since the optically-defined merging cluster catalog is unbiased against the cluster merger phase in contrast to X-ray and Sunyaev-Zel'dovich effect observations, X-ray follow-up study have the advantage to study cluster mergers at various phases.

We also searched for a radio relic or halo using the NRAO VLA Sky Survey data (NVSS; \cite{NVSS}). Although there is radio emission related to the radio galaxy (BCG), we could not find any diffuse radio emission. Since HSC J085024+001536 is a low-mass merger, it would be difficult to detect diffuse radio emissions because of low energy release \citep{Okabe2019}.

Numerical simulations \citep{2006ApJ...650..102A} studied the process of the cold front formation.
They found that the cold fronts generated by the first core passage can survive and continue to grow by disturbance of moving subhalos beyond the second core passage. 
The bottom center panel of figure 3 in \citet{2006ApJ...650..102A} shows multiple spiral structures of the sloshed gas and a cold front, similar to the case of HSC J085024+001536 (figure \ref{fig:img_residual}). The simulated image is at 2.8~Gyr after the first core passage, which agrees with our estimation. 

\citet{Simionescu12} found in the Perseus cluster that a large scale cold front exists at $\sim700$~kpc from the center, and \citet{Walker2018} estimated that its age is $\sim5$~Gyr. 
\citet{Rossetti2013} found a cold front at even larger scale (1~Mpc) in A2142.
These features are similar to the case of HSC J085024+001536.

\section{Summary}

We have carried out a joint X-ray, weak-lensing, and optical study of 
HSC J085024+001536 as the first target of the optically defined merging clusters.  

Both the galaxy density and the weak-lensing mass map showed the NW and SE peaks. 
The two-dimensional weak-lensing analysis revealed that the SE component is the main cluster with $M_{500}=1.35^{+0.76}_{-0.62}\times10^{14}h_{70}^{-1}M_\odot$ and the NW component is the sub cluster with $0.52^{+0.69}_{-0.36}\times10^{14}h_{70}^{-1}M_\odot$, respectively. The central position of each halo coincides with the brightest galaxy therein. 
We estimated the gas mass, $6.6^{+0.6}_{-0.5}\times10^{11}M_\odot$, and 
the weak-lensing mass, $1.7^{+1.4}_{-1.0}\times10^{13}M_\odot$, in a sphere of $1'$ radius from the NW peak. The resulting gas mas fraction, $f_{\mathrm{gas}}=4.0^{+2.3}_{-3.3}\%$, is  smaller than predicted by numerical simulation with AGNs \citep{2006ApJ...650..102A}. This is likely because the ICM of the NW peak had been stripped by ram pressure.

The extended X-ray emission around the SE peak has two peaks of which the east peak coincides with the BCG, and the west peak is the morphological center of the X-ray image. We simultaneously fitted the X-ray image with two $\beta$ models
and two point source models and subtracted the best-fit models from the X-ray image. The residual map clearly showed three arc-like excesses of which two are located at outer radii. The X-ray spectral analysis, by dividing the cluster region into four sectors, found the entropy changes across the ridges of the arc-like excesses. In particular, the north large-scale excess represents the temperature change $T_{\rm in}/T_{\rm out}=0.52^{+0.09}_{-0.11}$, which agrees with the typical feature of cold fronts. Combining with the ram-pressure stripping, the north cold excess is induced by the moving NW subhalo. The south excess is located at the symmetric position of the north excess  with respect to the morphological center. The tangentially elongated multiple excesses, their ages, and dynamical time suggests that 
HSC J085024+001536 is at a late phase merger, at least at the second impact or later, with a non-zero impact parameter.

\begin{ack}
This work based on observations obtained with XMM-Newton, an ESA science mission with instruments and contributions directly funded by ESA Member States and NASA.

\indent The Hyper Suprime-Cam (HSC) collaboration includes the astronomical communities of Japan and Taiwan, and Princeton University. The HSC instrumentation and software were developed by the National Astronomical Observatory of Japan (NAOJ), the Kavli Institute for the Physics and Mathematics of the Universe (Kavli IPMU), the University of Tokyo, the High Energy Accelerator Research Organization (KEK), the Academia Sinica Institute for Astronomy and Astrophysics in Taiwan (ASIAA), and Princeton University. Funding was contributed by the FIRST program from Japanese Cabinet Office, the Ministry of Education, Culture, Sports, Science and Technology (MEXT), the Japan Society for the Promotion of Science (JSPS), Japan Science and Technology Agency (JST), the Toray Science Foundation, NAOJ, Kavli IPMU, KEK, ASIAA, and Princeton University.

\indent This paper makes use of software developed for the Large Synoptic Survey Telescope. We thank the LSST Project for making their code available as free software at http://dm.lsst.org

\indent The Pan-STARRS1 Surveys (PS1) have been made possible through contributions of the Institute for Astronomy, the University of Hawaii, the Pan-STARRS Project Office, the Max-Planck Society and its participating institutes, the Max Planck Institute for Astronomy, Heidelberg and the Max Planck Institute for Extraterrestrial Physics, Garching, The Johns Hopkins University, Durham University, the University of Edinburgh, Queen’s University Belfast, the Harvard-Smithsonian Center for Astrophysics, the Las Cumbres Observatory Global Telescope Network Incorporated, the National Central University of Taiwan, the Space Telescope Science Institute, the National Aeronautics and Space Administration under Grant No. NNX08AR22G issued through the Planetary Science Division of the NASA Science Mission Directorate, the National Science Foundation under Grant No. AST-1238877, the University of Maryland, and Eotvos Lorand University (ELTE) and the Los Alamos National Laboratory.

\indent This work is supported in part by the Ministry of Science and Technology of Taiwan (grant MOST 106-2628-M-001-003-MY3) and by Academia Sinica (grant AS-IA-107-M01).

\indent This work is also supported in part by World Premier International Research Center Initiative (WPI Initiative), MEXT, Japan, and JSPS KAKENHI Grants Number JP15H05892, JP17H02868, JP18K03693 and JP19H05189.

\end{ack}

%
%

\bibliographystyle{pasj}
\bibliography{W535}


\appendix
\section*{Two-dimensional weak-lensing analysis}\label{sec:app}

We here describe the weak-lensing mass measurement using two-dimensional shear pattern.
The dimensional reduced shear $\Delta \Sigma_{\alpha}\, (\alpha=1,2)$ is computed by averaging the measured ellipticity $e_\alpha$ in the $k$-th grid box of $1.5'\times1.5'$;
\begin{eqnarray}
\Delta \Sigma_{\alpha} (\bm{x}_k) = 
\frac{\sum_{i} e_{\alpha,i} w_{i} \langle \Sigma_{{\rm cr}}(z_{l}, z_{s,i})^{-1}\rangle^{-1}}{2 \mathcal{R}({\bm x}_k) (1+K({\bm x}_k)) \sum_{i} w_{i}}, \label{eq:g+}
\end{eqnarray}
(e.g., \cite{Miyaoka18,Medezinski18b,Okabe2019,Miyatake19,Murata19,Umetsu2020,Okabe2019MUSTANG2}).
The two dimensional position, ${\bm x}_k$, is defined by the weighted harmonic mean \citep{Okabe16}.
The inverse of the mean critical surface mass density for the $i$-th galaxy is computed by the probability function $P(z)$ 
from the machine learning method (MLZ; \cite{MLZ14}) calibrated
with spectroscopic data \citep{HSCPhotoz17}.  
\begin{eqnarray}
 \langle \Sigma_{{\rm cr}}(z_{l},z_{s})^{-1}\rangle =
  \frac{\int^\infty_{z_{l}}\Sigma_{{\rm cr}}^{-1}(z_{l},z_{s})P(z_{s})dz_{s}}{\int^\infty_{0}P(z_{s})dz_{s}},
\end{eqnarray}
where $z_{l}$ and $z_{s}$ are the cluster and source redshift, respectively.
The critical surface mass density is $\Sigma_{{\rm cr}}=c^2D_{s}/4\pi G D_{l} D_{ls}$,
where $D_l$, $D_s$ and $D_{ls}$ are the angular diameter distances from
the observer to the cluster, to the sources and from the lens to the sources, respectively.
We use background galaxies satisfying $\int_{z_l+0.2} P(z_s)dz_s>0.98$ \citep{Medezinski18}. 

The log-likelihood is defined as 
\begin{eqnarray}
  -2\ln\mathcal{L}_{\rm WL} = \sum_{\alpha,
   \beta=1}^2\sum_{k,m}\left[\Delta \Sigma_{\alpha,k} - f_{{\rm model},\alpha}\left(\bm{R}_k\right) \right]\bm{C}^{-1}_{\alpha\beta, km}\nonumber \\
    \times\left[ \Delta \Sigma_{\beta,m} - f_{{\rm model},\beta}\left(\bm{R}_m\right) \right]+\ln(\det(C_{\alpha\beta,km})),
\end{eqnarray}
where the subscripts $\alpha$ and $\beta$ denote each shear component. $C_{\alpha\beta}$ denotes the covariance error matrix of shape measurement.
The central positions are restricted to full-width boxes $2\,{\rm arcmin} \times2\,{\rm arcmin}$ centered on the BCGs.
The two-dimensional analysis is good at determining the central
positions \citep{Oguri10b} and measuring masses of multi-components of
merging clusters \citep{Okabe11,Okabe15b,Medezinski16,Okabe2019MUSTANG2}. Since the concentration parameter cannot be constrained well, we assume the mass-concentration relation \citep{2015ApJ...799..108D}. 

\end{document}